\ifx\LaTeXe\undefined
 old LaTeX version
  \documentstyle[12pt,epsfig]{article}
  
\else
\documentclass[12pt]{article}
\usepackage{a4}
\usepackage[dvips]{graphicx}
\usepackage{latexsym}
\usepackage{euscript}

\fi

\def\simge{\mathrel{%   
   \rlap{\raise 0.511ex \hbox{$>$}}{\lower 0.511ex \hbox{$\sim$}}}}   
\def\simle{\mathrel{   
   \rlap{\raise 0.511ex \hbox{$<$}}{\lower 0.511ex \hbox{$\sim$}}}}   
\def\slashchar#1{\setbox0=hbox{$#1$}           % set a box for #1   
   \dimen0=\wd0                                 % and get its size   
   \setbox1=\hbox{/} \dimen1=\wd1               % get size of /   
   \ifdim\dimen0>\dimen1                        % #1 is bigger   
      \rlap{\hbox to \dimen0{\hfil/\hfil}}      % so center / in box   
      #1                                        % and print #1   
   \else                                        % / is bigger   
      \rlap{\hbox to \dimen1{\hfil$#1$\hfil}}   % so center #1   
      /                                         % and print /   
   \fi}                                         %   

\def\simge{\mathrel{%   
   \rlap{\raise 0.511ex \hbox{$>$}}{\lower 0.511ex \hbox{$\sim$}}}}   
\def\simle{\mathrel{   
   \rlap{\raise 0.511ex \hbox{$<$}}{\lower 0.511ex \hbox{$\sim$}}}}   
\def\slashchar#1{\setbox0=\hbox{$#1$}           % set a box for #1   
   \dimen0=\wd0                                 % and get its size   
   \setbox1=\hbox{/} \dimen1=\wd1               % get size of /   
   \ifdim\dimen0>\dimen1                        % #1 is bigger   
      \rlap{\hbox to \dimen0{\hfil/\hfil}}      % so center / in box   
      #1                                        % and print #1   
   \else                                        % / is bigger   
      \rlap{\hbox to \dimen1{\hfil$#1$\hfil}}   % so center #1   
      /                                         % and print /   
   \fi}
\newcommand{\pslash}{{\slashchar p}} 
 
\newcommand{\de}{\partial}
\newcommand{\Tr}{\mbox{Tr}\;} 
\newcommand{\Pj}{\mbox{I}\!\!\mbox{P}}
\newcommand{\id}{\mbox{1$\!\!$I}}

\newcommand{\RI}{\mbox{\scriptsize RI}}

\newcommand{\bc}{\begin{center}}
\newcommand{\ec}{\end{center}}
\newcommand{\be}{\begin{equation}}
\newcommand{\ee}{\end{equation}}
\newcommand{\bea}{\begin{eqnarray}}
\newcommand{\eea}{\end{eqnarray}}
\newcommand{\ba}{\begin{equation} \left\{ \begin{array}{lr}}
\newcommand{\ea}{\end{array} \right. \end{equation}}

\newcommand{\amu}{A_{\mu}}
\newcommand{\m}{_{\mu}}

\begin{document}
\pagestyle{empty} 
\vspace{-0.6in}
\begin{flushright}
CERN-TH/2002-079\\
CPT-2002/PE.4349\\
ROME1-1334/2002 \\
ROM2F/2002/11
\end{flushright}
\vskip 1.0 in

\centerline{\large {\bf{ Remarks on the Gauge Dependence}}}
\centerline{\large {\bf{ of the RI/MOM Renormalization Procedure}}}
\vskip 1.0cm
\centerline{L. Giusti$^{(a)}$\footnote{On leave of absence from Centre de Physique 
Th\'eorique, CNRS Luminy, Case 907, F-13288 Marseille Cedex 9, France}, 
S. Petrarca$^{(b,c)}$, B. Taglienti$^{(c)}$, N.~Tantalo$^{(d)}$}
\centerline{\small $^a$ Theory Division, CERN}
\centerline{CH-1211 Geneva 23, Switzerland}
\centerline{\small  $^b$ Dipartimento di Fisica, Universit\`a di Roma ``La Sapienza'',}
\centerline{\small P.le A. Moro 2, I-00185 Rome, Italy.}
\centerline{\small $^c$INFN, Sezione di Roma 1, P.le A. Moro 2, I-00185 Rome, Italy.}
\centerline{\small  $^d$ Dipartimento di Fisica, Universit\`a di Roma ``Tor Vergata'',}
\centerline{\small Via della Ricerca Scientifica 1, I-00133 Rome, Italy}

\centerline{\small }
\vskip 1.0in
\begin{abstract}
The RI/MOM non-perturbative renormalization scheme
is studied on the lattice in SU(3) 
quenched QCD with Wilson
fermions. The gauge dependence of some fermion
bilinear renormalization constants is discussed
by comparing data which have been gauge-fixed in two
different realizations of the Landau gauge and  
in a generic covariant gauge.
The very good agreement between the various sets of results
and the theory indicates that the numerical uncertainty
 induced by the lattice gauge-fixing 
procedure is moderate and below the statistical errors.
\end{abstract}
\vfill
\pagestyle{empty}\clearpage
\setcounter{page}{1}
\pagestyle{plain}
\newpage 
\pagestyle{plain} \setcounter{page}{1}

\newpage

\section{Introduction}
Non-perturbative renormalization techniques 
\cite{Martinelli:1994ty,luscher_np} have become a crucial 
ingredient in lattice computations of fundamental QCD parameters and hadronic 
matrix elements. They allow the limitations of lattice perturbation 
theory to be overcome
in the determinations of the relevant renormalization constants by
removing the uncertainties due to unknown higher order terms.
 
The RI/MOM technique proposed in Refs.~\cite{Martinelli:1994ty,Martinelli:dq}
has been successfully applied to compute renormalization constants
of composite fermion operators in many lattice regularizations:
Wilson fermions \cite{Martinelli:1994ty}, \cite{Martinelli:dq}--\cite{Martinelli:1997zc},
Kogut--Susskind fermions \cite{Aoki:1999mr}, domain-wall fermions \cite{Blum:2001sr}
and overlap fermions \cite{Giusti:2001pk}.
The method imposes RI/MOM renormalization conditions on conveniently defined 
amputated Green functions  computed 
non-perturbatively between off-shell quark states at large virtuality
in a fixed gauge. In practical computations the renormalization scale 
$\mu^2$, determined from the virtuality $p^2$ of the 
external states, must satisfy the 
condition $\Lambda_{\mbox{QCD}}\ll\mu\ll O(1/a)$ to ensure 
the reliability of the results \cite{Martinelli:1994ty}.

The RI/MOM renormalization constants 
depend in general on the external states and therefore on the gauge in which 
the RI/MOM renormalization conditions have been imposed. 
For gauge-invariant operators
the coefficients needed to match in a given gauge-invariant scheme will cancel these dependences 
up to higher orders in the continuum perturbative expansion and up to discretization errors.
A necessary assumption in this procedure is that, at large virtualities, the continuum 
limit of the 
renormalized gauge-dependent matrix elements, computed by imposing the non-perturbative 
Landau lattice gauge fixing, lead to the corresponding values computed 
in perturbation theory up to higher order \cite{Giusti:1999im}. 
The existence of both continuum \cite{Gribov} and lattice Gribov
copies \cite{Marinari:1991zv,defo91} and the numerical noise 
that they can generate \cite{pettassos}
(for a recent review see Ref.~\cite{gf_review}) 
is an unsolved problem of the 
lattice non-perturbative gauge fixing. The real concern is, of course, 
the influence that these phenomena may have on the values
of the renormalization constants, when computed using non-gauge-invariant 
quantities as in the RI/MOM scheme.

The aim of this paper is to study the systematics induced by the 
gauge-fixing procedure on the RI/MOM determinations of 
the renormalization constants. We have computed the quark field
($Z_\psi$), 
the axial-vector ($Z_A$) and scalar ($Z_S$) renormalization constants, 
by imposing the standard lattice Landau gauge 
and another realization of the
Landau condition that, in the following, will be called Landau1 gauge.
Whereas the two realizations impose the same gauge-fixing condition in the naive continuum
limit, they are affected by different Gribov ambiguities. 
By comparing the different sets of results for 
the renormalization constants, 
we have found negligible differences within our statistical errors.

As will be discussed in section 2, a very interesting feature of
the Landau1 gauge is that it can be generalized to
impose a generic covariant gauge on the lattice, as proposed in 
Ref.~\cite{Giusti:2001pk}. By exploiting this opportunity, 
we have extended the study of the 
gauge dependence to  off-shell Green functions
measured in a generic covariant gauge.

\section{Gauge-Fixing Conditions}\label{sec:CGF}
In this section we briefly describe the
continuum analogous of the non-perturbative gauge-fixing conditions
whose discretized forms have been used 
to obtain the numerical results presented below. 

In the standard Landau gauge the expectation value of a gauge-dependent 
operator is given by
\be
\langle {\cal O}_F \rangle = 
\int \delta\amu \, {\cal O}\, e^{-S(A)}
\Delta_{F}[A] \, \delta(\partial\m\amu) \; ,
\label{eq:expo}
\ee
where the Faddeev--Popov factor is defined as 
\be\label{eq:fadpop}
 \Delta_F[A] = 
\det{ 
\left(
\partial_\mu D_\mu[A]
\right)
}
\ee
and the gauge-fixing condition 
\be\label{eq:landau0}
\partial\m\amu^G = 0\; \mbox{with periodic boundary conditions}\, ,
\ee
is enforced non-perturbatively by minimizing 
the Gribov functional \cite{Gribov} 
\be\label{eq:landau}
F_A[G]\equiv ||A^G||^2=
\int d^4x\, \mbox{\rm Tr}\left[A^G_{\mu}A^G_{\mu}\right] \; .
\ee
Since the functional in Eq.~(\ref{eq:landau}) can have  
multiple minima in the non-perturbative regime, the 
corresponding gauge-fixing condition is incomplete, 
a fact that is known as the Gribov copy problem \cite{Gribov}.

In order to study the effects induced by Gribov's ambiguities
on gauge-dependent matrix elements, the Landau gauge can also be fixed
by minimizing a different functional \cite{Giusti:1999im}
\be\label{cov11}
H_A [G]\equiv 
\int d^4x\,\mbox{\rm Tr}\left[(\partial_{\mu}A^G_{\mu})^2\right]\; ,
\ee
which reaches its minima when 
\be\label{eq:brutta}
D_\nu \partial_\nu\partial_\mu A^G_\mu = 0 \, , 
\ee 
where $D_\nu$ is the covariant derivative.
Equation (\ref{eq:brutta}), supplemented with periodic boundary conditions, 
defines a new gauge which we call Landau1 in the following.
In this case the 
expectation value of a gauge-dependent operator is given by 
\be
\langle {\cal O}_H \rangle = 
\int \delta\amu \, {\cal O}\, e^{-S(A)}
\Delta_{H}[A] \, \delta(D_\nu \partial_\nu\partial\m\amu)\; .
\label{eq:expo2}
\ee
where $\Delta_{H}[A]$ is defined analogously as in Eq.~(\ref{eq:fadpop})
\cite{Giusti:1999im}. 
By comparing Eqs.~(\ref{eq:expo}) and (\ref{eq:expo2})
it is straightforward to show that the two
procedures lead to the same values of gauge-dependent matrix elements
in the perturbative regime \cite{nazario_tes}.

The solutions of Eq.~(\ref{eq:brutta}) are all absolute minima 
of $H_A[G]$, at variance with the copies associated with the 
standard functional in Eq.~(\ref{eq:landau}). Therefore
we expect a different distribution of Gribov's copies for the 
two procedures, which could generate different effects 
in off-shell matrix elements. It is clear that Eq.~(\ref{eq:brutta}) 
could also have  ``spurious'' solutions, 
which correspond to zero modes of the operator $D_\nu \partial_\nu$ 
and do not satisfy the gauge condition in Eq.~(\ref{eq:landau0}).
The numerical results of the exploratory study presented
in Ref.~\cite{Giusti:1999im}, and, a posteriori, the new 
results presented below, indicate that such a possibility 
can be discarded.\\ 

The functional $H_A[G]$ has been proposed in 
the literature \cite{Giusti:1996kf,Giusti:1999im} because it
can be easily generalized to the following form
\be
H_{(A,\Lambda)}[G]\equiv 
\int d^4x\, \mbox{\rm Tr}\left[(\partial_{\mu}A^G_{\mu}-\Lambda)  
(\partial_{\nu}A^G_{\nu}-\Lambda)\right]\; ,  
\label{cov111}
\ee
where $\Lambda(x)$ are matrices belonging to the Lie algebra of the
group and are generated according to a Gaussian distribution 
with the variance $\xi$ fixed at a given value. It is the value of this parameter 
that determines the corresponding covariant gauge. 
The $H_{(A,\Lambda)}[G]$ minima satisfy the gauge condition
\be
D_\nu \partial_\nu(\partial_\mu A^G_{\mu}-\Lambda) = 0 \;, 
\mbox{with periodic boundary conditions}\; .
\ee 
The expectation value of a given operator is obtained by 
\be
\langle {\cal O}_H \rangle_\xi =\int\delta\Lambda 
e^{-\frac{1}{\xi}\int d^4x \Tr [\Lambda^2]}
\int \delta\amu \, {\cal O}\, e^{-S(A)}
\Delta_H[A]
\delta(\partial\m\amu-\Lambda)\; .
\label{eq:o2}
\ee
For a complete description of the method, see Ref.~\cite{Giusti:1999im}.
In the following we perform a further check of the gauge-dependence 
comparing the results obtained in the Landau gauge
and in one of the lattice covariant gauges.

\section{Numerical Results and Discussion}\label{sec:NR}
We have used a sample of $80$ SU(3) gauge configurations retrieved from the
repository at the ``Gauge Connection'' (http://qcd.nersc.gov/), which 
were generated with the Wilson gluonic action at $\beta=6.0$
and $V \times T = 16^3 \times 32$ \cite{GaugeConn}. By using the discretized 
version (see~\cite{Giusti:1999im} for details) of the functionals 
in Eqs. (\ref{eq:landau}), (\ref{cov11}) and (\ref{cov111}) 
we have rotated each configuration in the Landau gauge, in the Landau1 gauge  and 
in the covariant gauge with the bare value of the gauge fixing parameter $\xi=8$
respectively. 

The various gauges have been enforced with a quality factor 
$\theta < 10^{-6}$ (see ~\cite{Giusti:1999im} for details)
with the Origin2000 computers at Boston University.
The inversion of the Wilson--Dirac operator has been 
performed on an INFN APE100 machine for masses
corresponding to the values of the $k$ parameter shown
in Table~\ref{tab:dettagliret}.
%--------------------------------------------------------------%
\begin{table}[hpbt]
\begin{center}
\begin{tabular}{|l|l|}
\hline\hline
$\beta$         &       6.0                     \\
\hline
Action          &       Wilson                  \\
\hline
Lattice Size    &       $16^3 \times 32$        \\
\hline
\# Configurations  &       80                      \\
\hline
                &       Landau                  \\
Gauges          &       Landau1                 \\
                &       $\xi=8$                 \\
\hline
                &       0.1530                  \\
                &       0.1535                  \\
$k$        &       0.1540                  \\
                &       0.1545                  \\
                &       0.1550                  \\
                &       0.1555                  \\
\hline
$k_c$~\cite{allton}      &       0.15703(2)              \\
\hline
$a^{-1}$~\cite{allton}        &       2.26(5) GeV           \\
\hline\hline
\end{tabular}
\end{center}
\caption{Main parameters of
our simulations.}
\label{tab:dettagliret}
\end{table}
%--------------------------------------------------------------%
Once the quark propagator $S(x,0) =  \langle \psi(x) \bar \psi(0) \rangle$ 
has been computed for each mass and gauge and Fourier-transformed, we have determined the  
inverse Euclidean bare-quark propagator $S^{-1}(pa)$.
We have computed the amputated Green functions of the local quark bilinears
$O_\Gamma(x)=\bar{\psi}(x) \Gamma \psi(x)$ in the Fourier 
space~\footnote{In what follows we adopt the conventions
of Ref.~\cite{Gimenez:1998ue}.} 
\begin{equation}
\Lambda_\Gamma(pa)=S^{-1}(pa)G_\Gamma(pa)S^{-1}(pa)\; ,
\end{equation}  
where
\be
G_\Gamma(p) =
\int{
d^4x d^4y \,e^{-ip(x-y)}\langle \psi(x) O_\Gamma(0) \bar{\psi}(y) \rangle} 
\ee
and $\Gamma$ is a generic Dirac matrix 
corresponding to 
$\Gamma = \left\{\mbox{A},\mbox{S} \right\} = \left\{\gamma_\mu\gamma_5,\id\right\}$.

In the RI/MOM scheme, the wave-function renormalization constant $Z_\psi$ is defined as 
\begin{equation}
Z_{\psi}(\mu a)=\lim_{m\rightarrow 0}-i
\frac{1}{12} \Tr \left[\frac{\de S^{-1}(pa)}{\de \pslash }\right]_{p^2=\mu^2}\; .
\label{eq:Z_q_WI} 
\end{equation}
In order to avoid numerical derivatives, 
it is convenient to use 
\begin{equation}
Z'_{\psi}(\mu a)=\lim_{m\rightarrow 0}\left.-i\frac{1}{12}\frac{\Tr \sum_{\mu=1,4}
\gamma_\mu \sin(p_\mu a)S^{-1}(pa)}
{4\sum_{\mu=1,4}\sin^2(p_\mu a)}\right|_{p^2=\mu^2}\; ,
\label{eq:Z_q'_WI} 
\end{equation}
which differs from $Z_{\psi}$ by a finite term of order
$\alpha_s$. The matching between the two
renormalization constants can be computed in continuum
perturbation theory. Since this correction is negligible
at the level of the numerical precision of our calculation (see below),
we will not take it into account in the following. 
Therefore throughout this paper we assume $Z_{\psi}=Z'_{\psi}$.
\begin{figure}[phtb]
\begin{center}
\includegraphics[width=15cm,height=15cm]{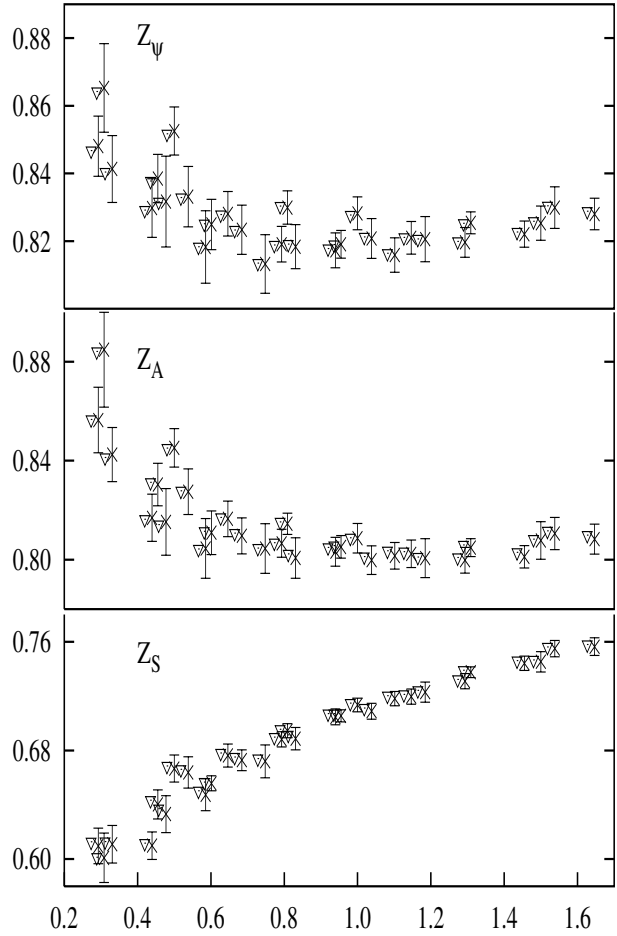}
\end{center}
\caption{The $Z_A$ and $Z_S$ renormalization constants are
shown as a function of $(a \mu)^2$ for the Landau and Landau1 gauges,
as discussed in the text.
The data have been slightly displaced in the $x$-direction and
the error bars for one set only have been shown in order
to help the eye.
The error bars for the other sets are comparable. The errors are jackknife.}
\label{fig:ssig1}
\end{figure}

The renormalization constant $Z_\Gamma^{\RI}(\mu a, g_0)$, which defines the 
renormalized operator $\hat {\cal O}_\Gamma^{\RI} =Z_\Gamma^{\RI} {\cal O}_\Gamma$,
is fixed by imposing in the chiral limit the renormalization condition
\begin{equation}
Z_\Gamma^{\RI}(\mu a)Z_{\psi}^{-1}(\mu a) \Tr\Pj_\Gamma\Lambda_\Gamma (pa)
|_{p^2=\mu^2}=1,
\label{eq:RI}
\end{equation}
where $\Pj_\Gamma$ is a suitable projector on the tree-level amputated
Green function \cite{Martinelli:1994ty}.\\

\begin{figure}[phtb]
\begin{center}
\includegraphics[width=15cm,height=8cm]{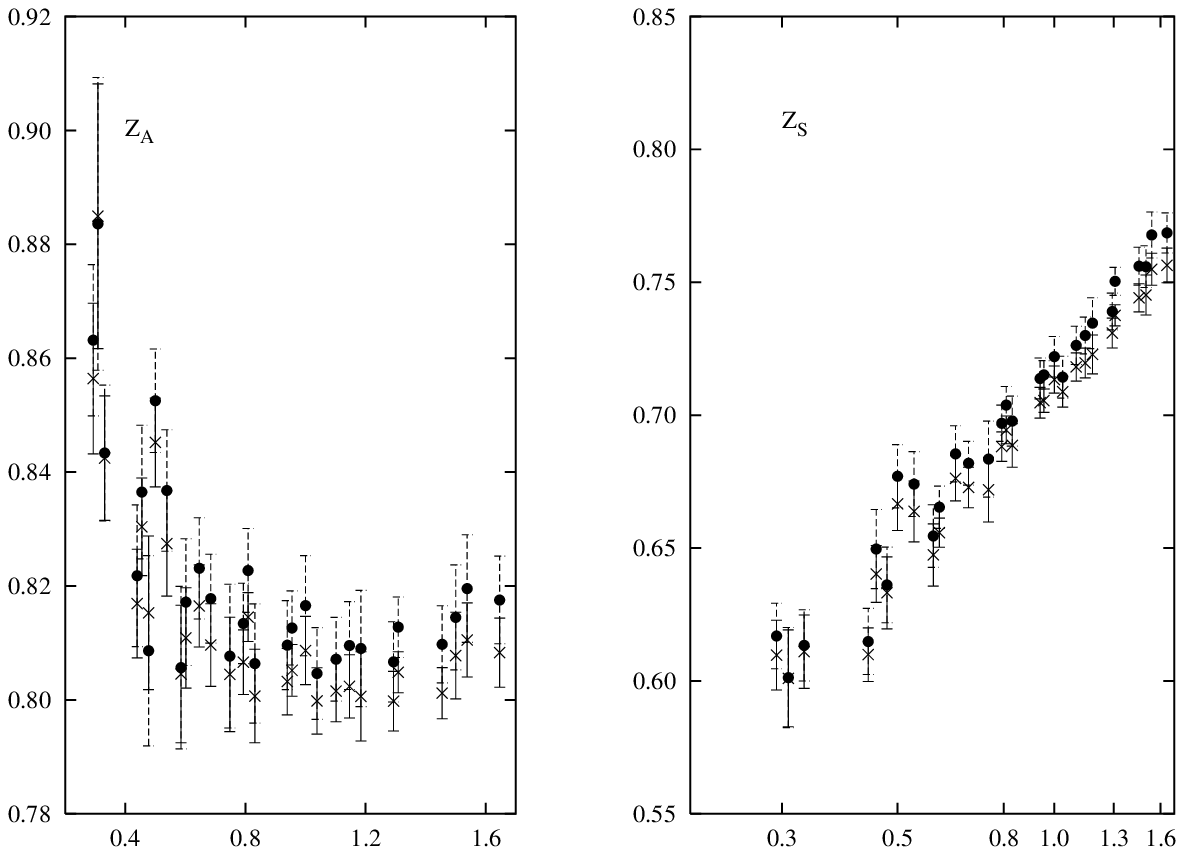}
\label{fig:ZaZs}
\end{center}
\caption{The behaviour of $Z_A$ and $Z_S$  is
shown as a function of $(a \mu)^2$ for the  Landau (black dots)
and the covariant gauge (crosses) data. The errors are jackknife.}
\end{figure}

Even though ${\cal O}_A$ and ${\cal O}_S$ are gauge-independent
operators, their matrix elements between quark states
acquire a gauge dependence. In Fig. {\ref{fig:ssig1}} are shown the numerical
results for $Z_\psi$, $Z_A$ and $Z_S$ calculated
in the Landau and in the Landau1 gauge, as a function of the square lattice momenta
\be
(ap_L^\mu)^2=4\sin^2\left(\frac{ap^\mu}{2}\right) .
\label{eq:latticemomenta}
\ee
In the perturbative region,
$Z_\psi(a \mu)$ has a flat dependence on the renormalization
scale. This is expected since
in the Landau gauge it has zero anomalous dimension 
at the leading order. Also $Z_A(a \mu)$ shows a flat behaviour 
due to the compatibility of the RI/MOM renormalization 
conditions with the axial Ward identities. The value of $Z_S(a \mu)$
increases with a logarithmic behaviour, as expected in
perturbation theory (see also Fig.~\ref{fig:ZaZs}).
Our data are in very good agreement with the results reported 
in the literature~\cite{Gimenez:1998ue}.

The results for all correlators, corresponding to Landau and Landau1 gauges,
coincide within the statistical errors. As a consequence we can conclude that 
the data do not show any residual effect due to the presence of 
lattice  Gribov's copies in the statistical sample of configurations
generated and for the lattice used. This represents one 
of the main results of this paper.
\begin{figure}[phtb]
\begin{center}
\includegraphics[width=15cm,height=15cm]{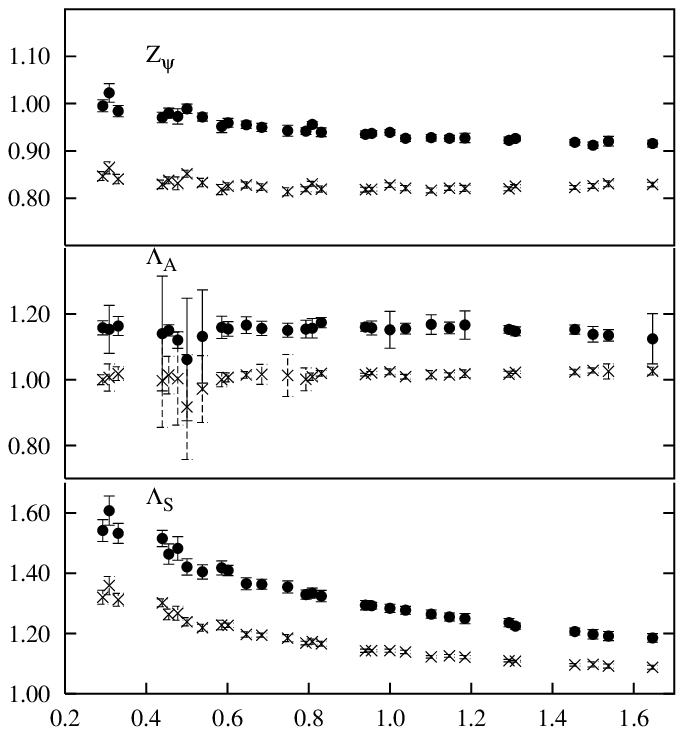}
\end{center}
\caption{The behaviour of $Z_\psi$, $\Lambda_A$ and $\Lambda_S$  is
shown as a function of $(a p)^2$ for the Landau (black dots)
and the covariant gauge (crosses) data. The errors are jackknife.}
\label{fig:ZpsiLaLs}
\end{figure}

To show the amount of gauge dependence that  can  be
found in our correlators and therefore our sensitiveness to the gauge, 
in the last part of this section we compare the results obtained 
in the Landau and in the covariant gauge with the bare 
gauge parameter fixed to $\xi=8$. 
The gauge dependence of the primary quantities 
$Z_\psi$, $\Lambda_A$ and $\Lambda_S$
measured on the lattice
is clearly visible in Fig.~\ref{fig:ZpsiLaLs}, where they
are plotted for the Landau and the covariant gauge as a function of 
the momenta $(ap)^2$. All three quantities show a statistically 
significant gauge dependence while the renormalization constants 
$Z_A$, $Z_S$ shown in Fig.~\ref{fig:ZaZs}, obtained by computing the ratios 
as indicated in Eq.~(\ref{eq:RI}),
do not exhibit any visible gauge dependence
within the statistical precision reached in our
simulation. For these quantities, the fluctuations of
the simulation hide the weak gauge 
dependence that is expected in
perturbation theory from the next-to-leading order 
terms in $\alpha_s$.
These results indicate an upper limit to the 
numerical troubles that can be expected in the
RI/MOM renormalization constants because
of the lattice gauge fixing. 

\section{Conclusions}\label{sec:C}
We have proposed a method to study the 
effects of Gribov's copies on lattice gauge-dependent matrix 
elements. We have implemented it numerically in quenched QCD
with Wilson fermions and we have computed
the matrix elements needed to fix the quark field, 
the axial-vector and scalar renormalization constants
in the RI/MOM non-perturbative renormalization technique.
Our numerical results do not show any significant
differences between data coming from the 
standard Landau and the Landau1 gauge. We 
do not find any relevant effect that can 
be traced back to the incompleteness of the gauge fixing
for a number of gauge-field configurations
and for lattice sizes of the order of those
used in this calculation.
The main indication of our paper is therefore that,
at the level of the precision that we have reached, 
the systematics due to the numerical gauge-fixing machinery
has negligible effect on the matrix elements needed 
for implementing  the RI/MOM scheme.
  
\section*{Acknowledgements}
We warmly thank N. Cabibbo, M.~L. Paciello,~F. Rapuano,
C. Rebbi, G.C. Rossi and M. Testa for many fruitful discussions 
on this subject. We thank the Center for Computational Science of Boston 
University, where part of the numerical computation has been done.
We also gratefully acknowledge the use of the gauge configurations 
produced by the authors in Ref.~\cite{GaugeConn}.
This work was partially supported  by the M.U.R.S.T. and
 by the European Community's 
Human Potential Programme under contract HPRN-CT-2000-00145, 
Hadrons/Lattice QCD.

\end{document}